\documentclass[12pt,a4paper]{article}

\usepackage{epsfig}
\usepackage{amsmath,amsfonts,amssymb}
\usepackage{t1enc}

\providecommand{\openone}{\leavevmode\hbox{\small1\kern-3.8pt\normalsize1}}

\newcommand{\vtb}{V_{tb}}
\newcommand{\vts}{V_{ts}}
\newcommand{\vtd}{V_{td}}
\newcommand{\fbin}{fb$^{-1}$}
\newcommand{\gm}{\gamma^\mu}
\newcommand{\smn}{\sigma^{\mu \nu}}

\newcommand{\ptmiss}{p_T\!\!\!\!\!\!\!\!\not\,\,\,\,\,\,\,}

\begin{document}

\begin{center}
\begin{Large}
{\bf Using single top rapidity to measure $V_{td}$, $V_{ts}$, $V_{tb}$ \\[2mm]
at hadron colliders}
\end{Large}

\vspace{0.5cm}
J. A. Aguilar--Saavedra$^a$, A. Onofre$^b$ \\[0.2cm] 
{\it $^a$ Departamento de Física Teórica y del Cosmos and CAFPE, \\
Universidad de Granada, E-18071 Granada, Spain} \\[0.1cm]
{\it $^b$ Departamento de Fisica, Universidade do Minho, P-4710-057 Braga, Portugal} \\[0.1cm]
\end{center}

\begin{abstract}
Single top production processes are usually regarded as the ones in which $\vtb$ can be directly measured at hadron colliders. We show that the analysis of the single top rapidity distribution in $t$-channel and $tW$ production can also set direct limits on $\vtd$. At LHC with 10 \fbin\ at 14 TeV the combined limits on $\vtd$ may be reduced by almost a factor of two when the top rapidity distribution is used. This also implies that the limits on $\vtb$ can also be reduced by 15\%, since both parameters as well as $\vts$ must be simultaneously obtained from a global fit to data.
At Tevatron the explotation of this distribution would require very high  statistics.
\end{abstract}

\section{Introduction}

In the Cabibbo-Kobayashi-Maskawa (CKM) matrix \cite{Cabibbo:1963yz,Kobayashi:1973fv} describing quark mixing, the matrix elements $\vtd$, $\vts$, $\vtb$ 
in the third row are the ones for which direct measurements are less precise.  Yet, the determination of these mixing parameters is of the utmost importance, in particular to test the CKM description of the observed CP violation in the $K$ and $B$ meson systems (see for example Ref.~\cite{Battaglia:2003in} and references there in). Within the Standard Model (SM), $\vtd \simeq 0.009$, $\vts \simeq 0.04$, $\vtb \simeq 1$, but substantial deviations from these predictions, based on $3 \times 3$ CKM unitarity, are possible in SM extensions. For example, the mixing of the top quark with a heavy charge $2/3$ quark isosinglet allows for $\vtb$ significantly smaller than unity~\cite{AguilarSaavedra:2002kr} while the mixing with a hypercharge $-1/3$ quark triplet may result in $\vtb > 1$~\cite{delAguila:2000aa} in sharp contrast with the SM unitarity bound $|\vtb|^2 \leq 1$.

Several collider observables can probe the top mixing with SM quarks.
Top pair production can measure the ratio
\begin{equation}
R = \frac{\mathrm{Br}(t \to Wb)}{\mathrm{Br}(t \to Wq)} = \frac{|\vtb|^2}{|\vtd|^2+|\vts|^2+|\vtb|^2}
\label{ec:R}
\end{equation}
(with $q=d,s,b$),
by comparing event samples with zero, one and two $b$ tags. Recently, the possibility of $s$ tagging has been explored~\cite{Ali:2010xx}, which would yield a measurement of
\begin{equation}
R' = \frac{\mathrm{Br}(t \to Ws)}{\mathrm{Br}(t \to Wb)} = \frac{|\vts|^2}{|\vtb|^2}\,,
\label{ec:Rp}
\end{equation}
by comparing the number of events with $b,s$ tags and with two $b$ tags.
Single top production processes have total cross sections which can be generically written as
\begin{equation}
\sigma = A_d |\vtd|^2 + A_s |\vts|^2 + A_b |\vtb|^2 \,,
\label{ec:xsec}
\end{equation}
with $A_{d,s,b}$ numerical constants (see the next section). But clearly, the ratios $R,R'$ and the several single top and antitop cross sections do not exhaust all possible observables sensitive to $\vtd$, $\vts$ and $\vtb$. In this paper we will show that the rapidity distribution of single top quarks is a very good discriminant between initial states with $d$  valence quarks against $s$ and $b$. Therefore, the inclusion of this observable in global fits allows to obtain much better constraints on $\vtd$, which also translate into more stringent bounds on $\vtb$, once that the three top CKM mixings must be simultaneously obtained from the fit. Besides, we note that rapidity analyses are well known for the determination of $Z'$ boson couplings to quarks~\cite{delAguila:1993ym}
 but have been rarely used in top physics.

In the following section we review the constraints that $R$ and the different single (anti)top cross sections place on the ($\vtd,\vts,\vtb$) parameter space, extending previous work in Ref.~\cite{Alwall:2006bx} to the Large Hadron Collider (LHC) for which single top production has many different features from Tevatron.
In section~\ref{sec:3} we discuss the top rapidity distributions and their uncertainties, including a brief analysis of the experimental issues regarding the top rapidity measurement. In section~\ref{sec:3b} we incorporate the rapidity distributions into global fits to
$\vtd$, $\vts$ and $\vtb$ for LHC and Tevatron, showing how they may improve the determination of $\vtd$ and $\vtb$. We point out that the full explotation of the top rapidity distribution, as any other precision analysis, requires sufficient statistics and excellent knowledge of the SM backgrounds. For this reason we limit ourselves to LHC at 14 TeV with 10 \fbin, giving for completeness results for Tevatron. We summarise our results in section~\ref{sec:4}.

\section{Constraints from cross sections and $R$}

\begin{figure}[htb] 
\begin{center}
\begin{tabular}{ccccc}
\epsfig{file=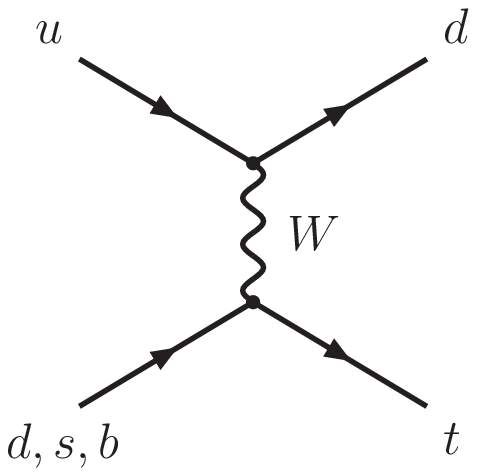,height=2.5cm,clip=} & \quad \quad
\epsfig{file=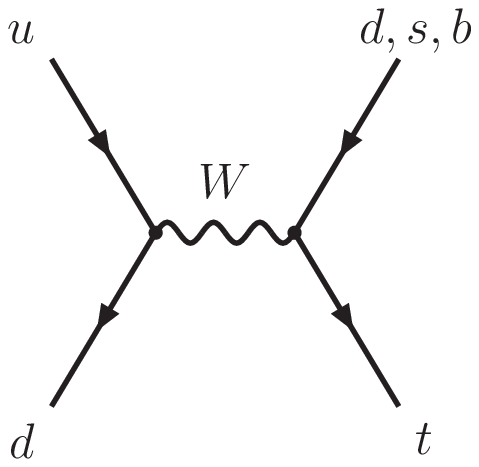,height=2.5cm,clip=} & \quad \quad
\epsfig{file=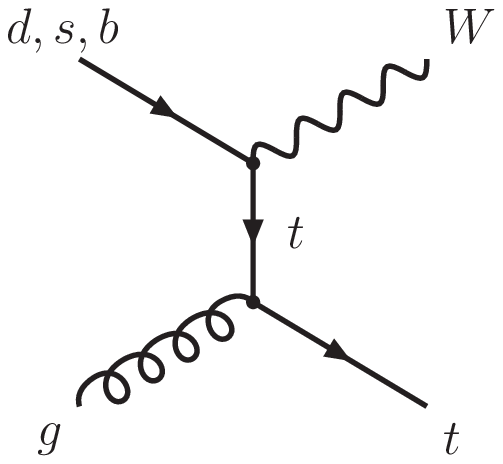,height=2.5cm,clip=} 
\end{tabular}
\caption{Representative Feynman diagrams for single top production in the $t$-channel (left), $s$-channel (center) and $tW$ processes (right).}
\label{fig:diags}
\end{center}
\end{figure}

There are three single top production processes at hadron colliders, usually denoted as $t$-channel (also abbreviated here as $tj$), $s$-channel (also $t \bar b$) and $tW$ production. Representative Feynman diagrams for these processes are depicted in Fig.~\ref{fig:diags}, including top quark mixing with the down-type quarks $d,s,b$.
For $t$-channel and $tW$ production the final state is in all cases the same (a top quark plus a jet or a $W$ boson) and the total cross sections have the form in Eq.~(\ref{ec:xsec}) with $A_d > A_s > A_b$ because of the larger parton distribution functions (PDFs) for $d$ and $s$ initial states. If the produced top quarks are reconstructed in the decay $t \to Wb$ with a tagged $b$ jet (as it is likely to happen in present and future analyses) the single top cross sections must also include an extra $R$ factor as in Eq.~(\ref{ec:R}) to take into account the branching ratio into $b$ quarks. 
For $s$-channel production the final state is a $b$ or a light quark. Dedicated searches for this process require two $b$ tags in order to distinguish it from $t$-channel production (as well as other selection criteria, as for example the absence of energetic forward jets). Hence, the only contribution to the measured cross section results from the $|\vtb|^2$ term in Eq.~(\ref{ec:xsec}), multiplied by a $R$ factor. On the other hand, $s$-channel $t \bar d$ and $t \bar s$ production will contribute to the $tj$ final state but the extra jet is more central than in the $t$-channel process, and depending on the particular event selection criteria these extra contributions may be highly suppressed. In the rest of this section we discuss the results for LHC and Tevatron in turn.

\subsection{Constraints at LHC}

At LHC (with a centre of mass energy of 14 TeV) the tree-level single top and antitop cross sections, including the branching ratio for $t \to Wb$, are
\begin{eqnarray}
\sigma(tj) & = & \left[ 678.6 \,|\vtd|^2 + 270.2 \,|\vts|^2 + 149.1 \,|\vtb|^2 \right] R ~\text{pb} \,, \notag \\
\sigma(\bar tj) & = & \left[ 233.3 \,|\vtd|^2 + 163.0 \,|\vts|^2 + 84.17 \,|\vtb|^2 \right] R ~\text{pb} \,, \notag \\
\sigma(t \bar b) & = & 4.28 \,|\vtb|^2 R ~\text{pb} \,, \notag \\
\sigma(\bar t b) & = & 2.61 \,|\vtb|^2 R ~\text{pb} \,, \notag \\
\sigma(tW) & = & \left[ 259.4 \,|\vtd|^2 + 59.78 \,|\vts|^2 + 27.57 \,|\vtb|^2 \right] R ~\text{pb} \,, \notag \\
\sigma(\bar tW) & = & \left[ 94.81 \,|\vtd|^2 + 59.78 \,|\vts|^2 + 27.57 \,|\vtb|^2 \right] R ~\text{pb} \,.
\label{ec:xsec-lhc}
\end{eqnarray}
They have been obtained with {\tt Protos}~\cite{AguilarSaavedra:2008gt} using CTEQ6L1 PDFs~\cite{Pumplin:2002vw} and setting $m_t = 175$ GeV. 
For simplicity, we will assume in this work that the charged current vertices have the SM structure, with a left-handed $\gm$ coupling. We note, however, that the most general gauge boson vertices also include right-handed $\gm$ as well $\smn$ terms~\cite{AguilarSaavedra:2008zc}. These anomalous contributions,
expected to be most important for the $Wtb$ vertex \cite{Bernreuther:2008us},
can also be included in the fit in a straightforward way by extending the set of observables~\cite{AguilarSaavedra:2006fy}. We also ignore possible new physics contributions from four-fermion operators~\cite{AguilarSaavedra:2010zi}.

We assume the following sensitivities for cross section (top plus antitop) measurements with 10 \fbin~\cite{Aad:2009wy},
\begin{align}
& \text{$t$-channel}:
 && \frac{\Delta \sigma}{\sigma} = 1.8\% ~\text{(stat)} \oplus 10\% ~\text{(sys)} \,, \notag \\[1mm]
& \text{$s$-channel}:
 && \frac{\Delta \sigma}{\sigma} = 20\% ~\text{(stat)} \oplus 48\% ~\text{(sys)} \,, \notag \\[1mm]
& tW:
 && \frac{\Delta \sigma}{\sigma} = 6.6\% ~\text{(stat)} \oplus 19.4\% ~\text{(sys)} \,.
\label{ec:statLHC}
\end{align}
For separate $t$, $\bar t$ measurements we rescale the statistical uncertainties above by the expected number of $t$, $\bar t$ events. In the fits we also include for completeness the theoretical uncertainties on cross sections, which are much smaller than the experimental ones and have negligible impact on our results: 3\% and 4\% for $t$-channel ($t$ and $\bar t$, respectively) \cite{Campbell:2009ss}, 6\% for $s$-channel \cite{Sullivan:2004ie} and 4.4\% for $tW$~\cite{Campbell:2005bb}.
We remark that these are uncertainties in the total rates, not in the distribution shapes (the uncertainties in the rapidity distributions are discussed in the next section). For this reason we conservatively take their numerical values at next-to-leading order (NLO) rather than at leading order (LO), which are larger. By taking smaller uncertainties in the total rates, the improvements found by using the rapidity analysis, insensitive to these but with uncertainties dominated by statistics, is conservatively smaller.
This procedure also seems adequate bearing in mind that in a first approximation the effect of NLO corrections can be accounted for by a global factor multiplying the cross sections. For $R$ we assume a precision
\begin{equation}
\Delta R = 0.5\% ~\text{(stat)} \oplus 5\% ~\text{(sys)} \,,
\end{equation}
with the same luminosity, extrapolating results in Ref.~\cite{Volpe:2009ns} and assumming an eventual improvement of systematic errors with larger data samples.

\begin{figure}[htb]
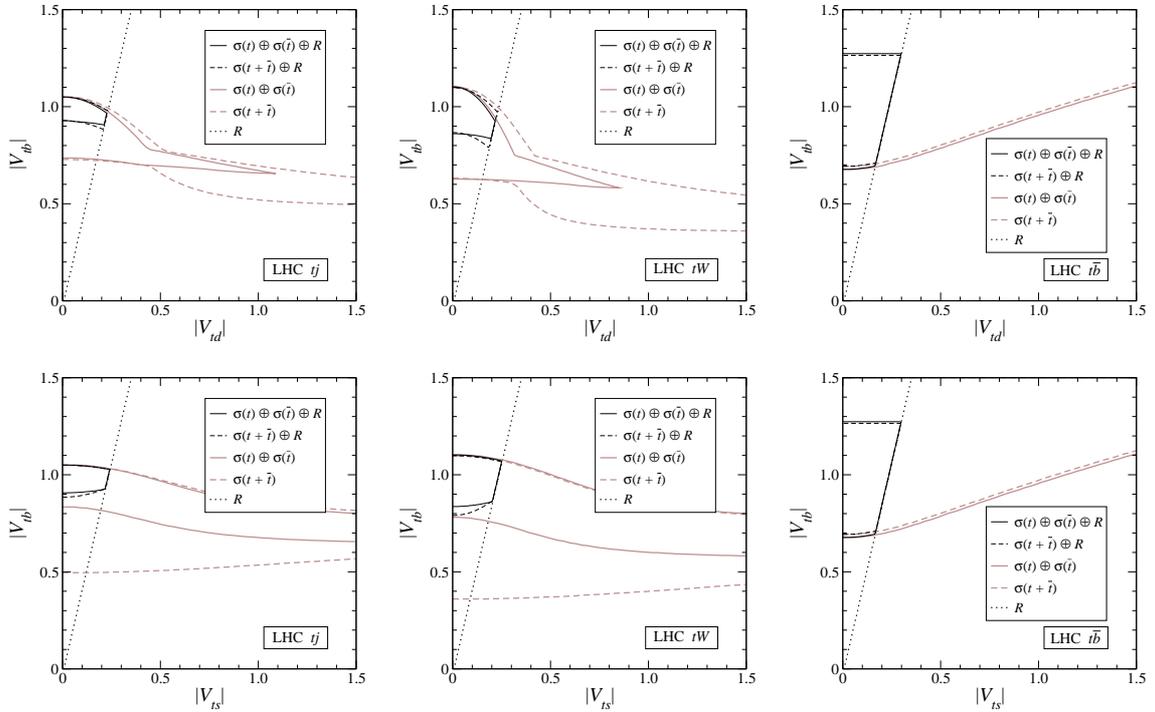
 
\begin{center}
\begin{tabular}{ccc}
\epsfig{file=Figs/fig2a.eps,height=4.5cm,clip=} &
\epsfig{file=Figs/fig2b.eps,height=4.5cm,clip=} &
\epsfig{file=Figs/fig2c.eps,height=4.5cm,clip=} \\[2mm]
\epsfig{file=Figs/fig2d.eps,height=4.5cm,clip=} &
\epsfig{file=Figs/fig2e.eps,height=4.5cm,clip=} &
\epsfig{file=Figs/fig2f.eps,height=4.5cm,clip=} \\[2mm]
\end{tabular}
\caption{Projections of the limits from single top cross section measurements
on the $(\vtd,\vtb)$ and $(\vts,\vtb)$ planes (gray lines). The solid and dashed black lines show the limits in combination with $R$. The black dotted lines show the limits from $R$ alone.}
\label{fig:regL1}
\end{center}
\end{figure}

In order to understand how the different processes constrain the ($\vtd,\vts,\vtb$) parameter space, we show in Fig.~\ref{fig:regL1} the $1\sigma$ limits on the mixings set by single top cross section measurements, either summing top plus antitop or separating them.
In the latter case we require that {\em both} $t$ and $\bar t$ cross sections are within a $1\sigma$ interval from the SM prediction.\footnote{For a better illustration of the interplay among the different observables, in the combined limits shown in this section we require that each of the observables is within $\pm 1\sigma$ of its SM prediction. This is different from requiring $1\sigma$ in the global fit to all observables, as it is done in the next section.}
The projection on the ($\vtd,\vts$) plane is not shown because
$\vtd$ and $\vts$ are unconstrained. (In the plots shown, they are left to vary in the interval $[0,1.5]$.)
As we can see, the different functional dependence on the three mixings of top and antitop cross sections can be exploited to improve the constraints by separating events with a top quark (giving a positive charge lepton) from those with an antiquark (with a negative charge). Noticeably, the constraints on $\vtd$ from $tW$ combined with $\bar tW$ are more stringent than from $tj$ combined with $\bar tj$, due to the more pronounced differences between $t$ and $\bar t$ production for the former.
We also observe that $t$-channel and $tW$ production give similar limits in parameter space, while those from $s$-channel are complementary. 

\begin{figure}[htb]
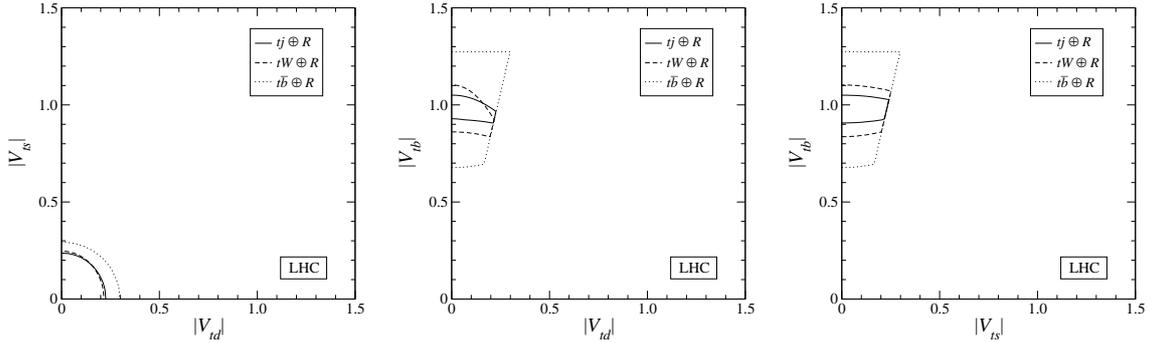

\begin{center}
\begin{tabular}{ccc}
\epsfig{file=Figs/fig3a.eps,height=4.5cm,clip=} &
\epsfig{file=Figs/fig3b.eps,height=4.5cm,clip=} &
\epsfig{file=Figs/fig3c.eps,height=4.5cm,clip=} \\[2mm]
\end{tabular}
\caption{Projections on the $(\vtd,\vts)$, $(\vtd,\vtb)$ and $(\vts,\vtb)$ planes of the limits from each single top process in combination with $R$.}
\label{fig:regL2}
\end{center}
\end{figure}

The limits from each single top process in combination with $R$ are also shown in each plot. We point out the apparent paradox that the two-dimensional combined limits are much smaller than the overlap of the different areas. This is easily understood because the allowed regions are actually three-dimensional volumes and the areas shown in the plots are their projection in different planes.
For a better comparison we also present in Fig.~\ref{fig:regL2} the limits from each single top process combined with $R$. Clearly, if the $R$ measurement is taken into account in the fits, $s$-channel production does not give any extra constraint, even if systematic uncertainties were reduced by a factor of two. We also observe that the limits from $t$-channel are more restrictive than those from $tW$ except for a small region in the $(\vtd,\vtb)$ plane.

From this analysis we can conclude that limits can be set on the mixings by combining at least 
\begin{itemize}
\item $t$-channel or $tW$ production plus $R$,
\item $t$-channel or $tW$ production plus $s$-channel,
\item $s$-channel plus $R$
\end{itemize}
although the first ones provide the most stringent constraints at LHC, given the large experimental uncertainty for $s$-channel cross section measurements.

\subsection{Limits at Tevatron}

At Tevatron the tree-level single top plus antitop cross sections, including the branching ratio for $t \to Wb$, are
\begin{eqnarray}
\sigma(tj+\bar tj) & = & \left[ 20.72 \,|\vtd|^2 + 5.476 \,|\vts|^2 + 1.838 \,|\vtb|^2 \right] R ~\text{pb} \,, \notag \\
\sigma(t \bar b+\bar tb) & = & 0.5245 \,|\vtb|^2 R ~\text{pb} \,,
\label{ec:xsec-tev}
\end{eqnarray}
obtained using {\tt Protos} with CTEQ6L1 PDFs. The theoretical uncertainty is taken as 9.3\% for $t$-channel \cite{Campbell:2009ss} and 13\% for $s$-channel \cite{Sullivan:2004ie}. As it is well known, $tW$ production has a very small cross section to be measured, and only amounts to a small correction to the $tj$ final state.\footnote{For initial $d$ and $s$ quarks its cross section is also much smaller than for $tj$ production, too, and its contribution would be further suppressed by the event selection criteria designed for the $t$-channel kinematics.} 
We assume the sensitivity
\begin{align}
& s+t~\text{channels}:
 && \frac{\Delta \sigma}{\sigma} = 10\% ~\text{(stat)} \oplus 18.5\% ~\text{(sys)} \,,
\end{align}
extrapolating the statistical uncertainty in the CDF analysis of Ref.~\cite{BDT} to 12 \fbin\ and assumming an eventual reduction of the systematic uncertainty to $3/4$ of its present value.
For separate $s$- and $t$-channel measurements we rescale the statistical uncertainties above by the SM cross sections corresponding to each process, and keep the same systematics.
For $R$ we take the measured value $R = 0.97^{+0.09}_{-0.08}$
\cite{Abazov:2008yn}. 

\begin{figure}[htb]
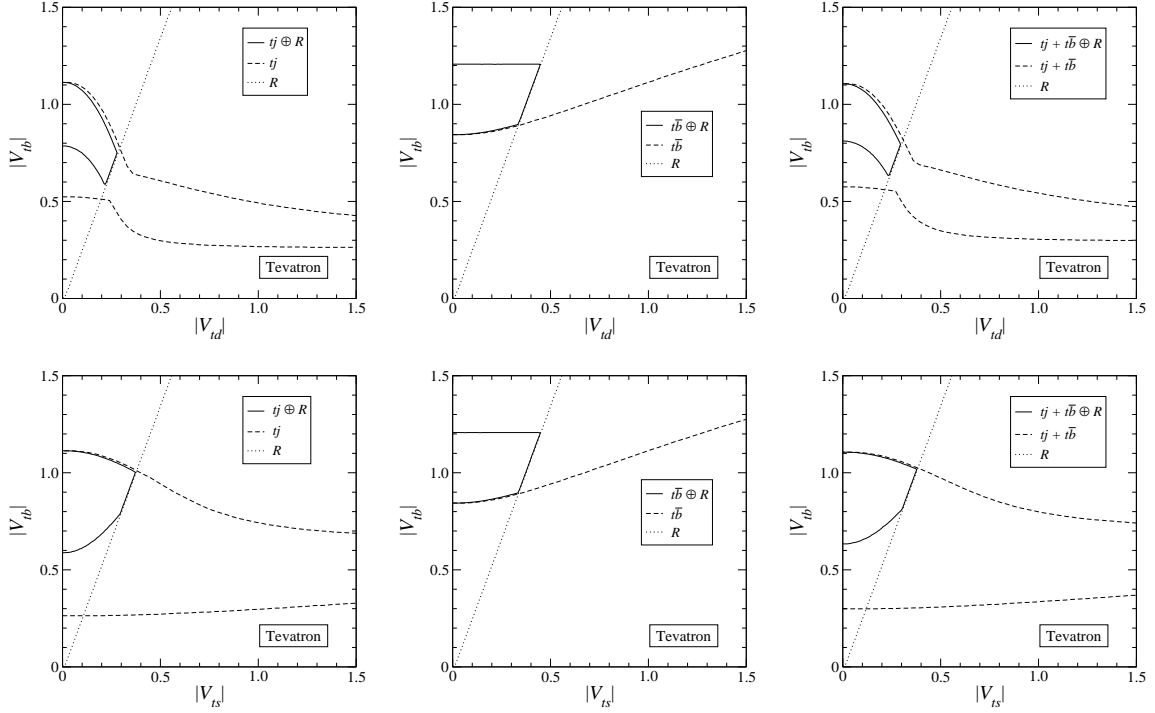

\begin{center}
\begin{tabular}{ccc}
\epsfig{file=Figs/fig4a.eps,height=4.5cm,clip=} &
\epsfig{file=Figs/fig4b.eps,height=4.5cm,clip=} &
\epsfig{file=Figs/fig4c.eps,height=4.5cm,clip=} \\[2mm]
\epsfig{file=Figs/fig4d.eps,height=4.5cm,clip=} &
\epsfig{file=Figs/fig4e.eps,height=4.5cm,clip=} &
\epsfig{file=Figs/fig4f.eps,height=4.5cm,clip=} \\[2mm]
\end{tabular}
\caption{Projections of the limits from single top cross section measurements
on the $(\vtd,\vtb)$ and $(\vts,\vtb)$ planes (gray lines). The solid and dashed black lines show the limits in combination with $R$. The black dotted lines show the limits from $R$ alone.}
\label{fig:regT1}
\end{center}
\end{figure}

We show in Fig.~\ref{fig:regT1} the limits on the mixings set by single top cross section measurements, either separating $s$- and $t$-channel production or summing both. We observe that the latter case is practically equivalent to measuring the $t$-channel cross section alone. If $s$- and $t$-channel cross sections are measured independently, the constraints they set are complementary. We also show in each plot the constraints from single top cross sections in combination with the measurement of $R$. As in the previous subsection, when combining observables we require that each of them is within $1\sigma$ of its SM value.

\begin{figure}[htb]
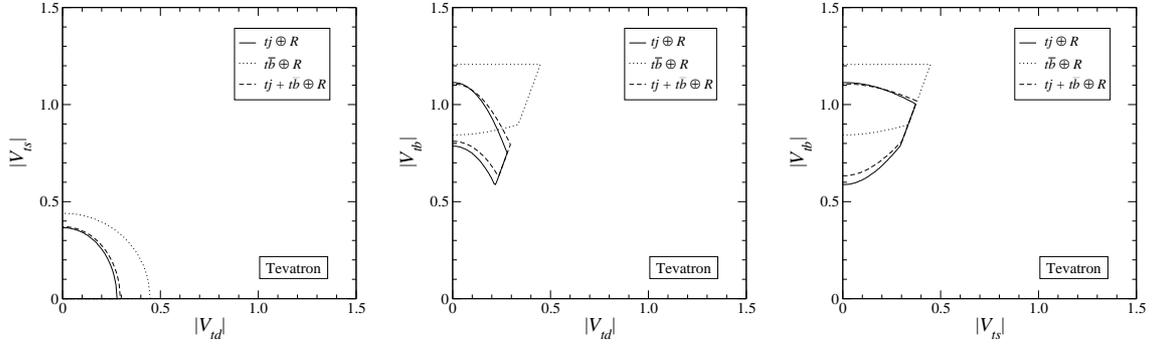

\begin{center}
\begin{tabular}{ccc}
\epsfig{file=Figs/fig5a.eps,height=4.5cm,clip=} &
\epsfig{file=Figs/fig5b.eps,height=4.5cm,clip=} &
\epsfig{file=Figs/fig5c.eps,height=4.5cm,clip=} \\[2mm]
\end{tabular}
\caption{Projections on the $(\vtd,\vts)$, $(\vtd,\vtb)$ and $(\vts,\vtb)$ planes of the limits from each single top process in combination with $R$.}
\label{fig:regT2}
\end{center}
\end{figure}

The complementarity of the different measurements can also be observed in Fig.~\ref{fig:regT2}, where we simultaneously present the constraints from each process and their sum, combined with $R$. Setting limits on the top CKM mixings requires at least
\begin{itemize}
\item $s$-channel plus $t$-channel,
\item $t$-channel or $s$-channel plus $R$,
\item $s+t$ channels plus $R$.
\end{itemize}
As pointed out in Ref.~\cite{Alwall:2006bx}, the measurement of the $s$-channel cross section at Tevatron is of great help in setting constraints on the top CKM matrix elements. Still, useful limits can be set even if the $s$- and $t$-channel cross sections cannot be measured independently with a good precision, as long as the measurement is combined with $R$.

\section{Single top rapidity distributions}
\label{sec:3}

\begin{figure}[htb]
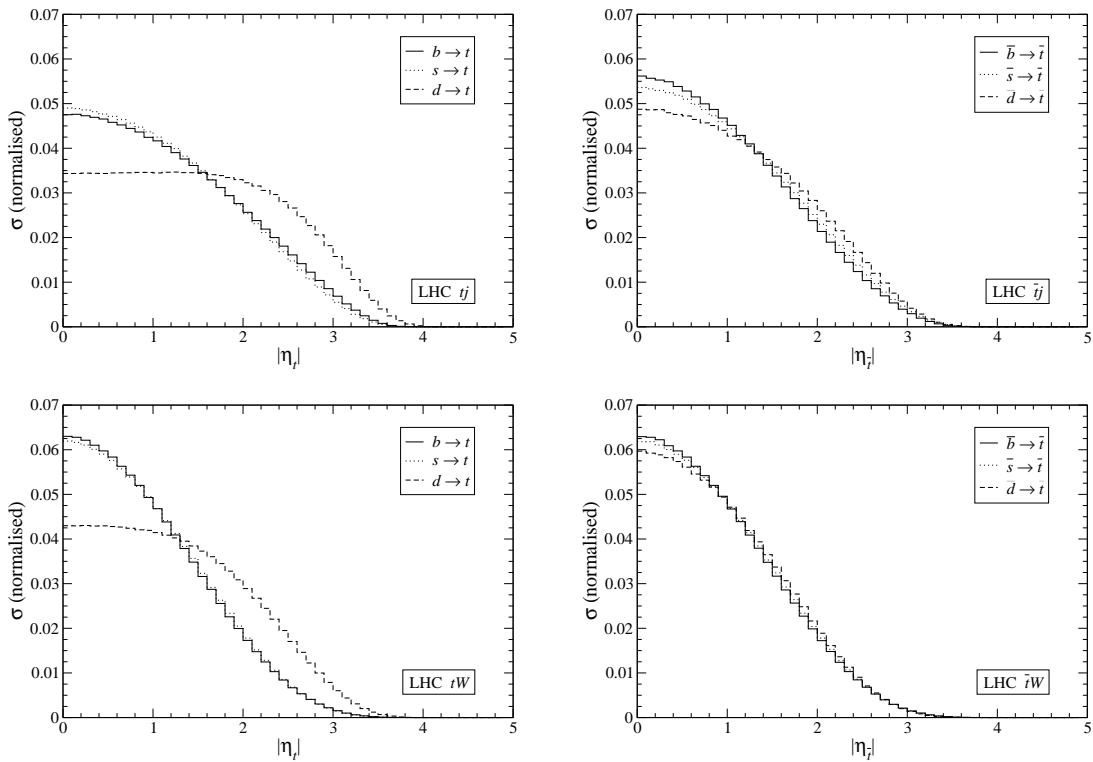
 
\begin{center}
\begin{tabular}{ccc}
\epsfig{file=Figs/fig6a.eps,height=4.8cm,clip=} & \quad
\epsfig{file=Figs/fig6b.eps,height=4.8cm,clip=} \\[2mm]
\epsfig{file=Figs/fig6c.eps,height=4.8cm,clip=} & \quad
\epsfig{file=Figs/fig6d.eps,height=4.8cm,clip=}
\end{tabular}
\caption{Normalised rapidity distributions for $t$-channel and $tW$ single top production at LHC.}
\label{fig:rapLHC}
\end{center}
\end{figure}

\begin{figure}[htb] 
\begin{center}
\epsfig{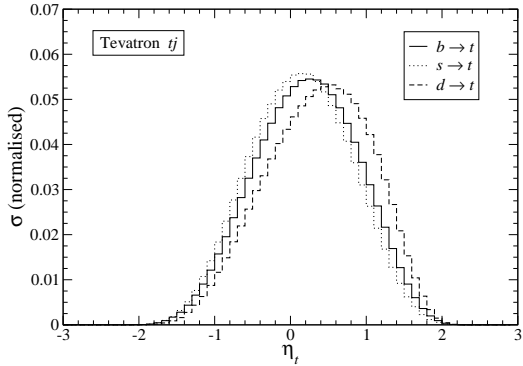}
\caption{Normalised rapidity distribution for $t$-channel single top production at Tevatron.}
\label{fig:rapTEV}
\end{center}
\end{figure}

In $t$-channel and $tW$ production from initial $s$, $b$ sea quarks, the events are more central than those resulting from initial $d$ quarks. We present in Fig.~\ref{fig:rapLHC} the normalised rapidity distributions for LHC at LO, calculated with {\tt Protos}. For $t$-channel production we use $Q^2=-p_W^2$ (with $p_W^2 < 0$) for the light quark line and $Q^2=-p_W^2+m_t^2$ for the one with the top quark. For $tW$ production we set $Q=m_t+M_W$.
For top quarks (left) the differences are quite remarkable, while the distributions for antiquarks are more similar. This fact makes even more important the separation between $t$ and $\bar t$ production in experimental analyses. In Fig.~\ref{fig:rapTEV} we present the rapidity distributions for $t$-channel production at Tevatron, with the positive $z$ axis chosen as the direction of the proton beam. Although the differences are not as significative as for LHC, they could still improve the constraints obtained from cross section measurements alone for high luminosities.

\begin{figure}[htb] 
\begin{center}
\epsfig{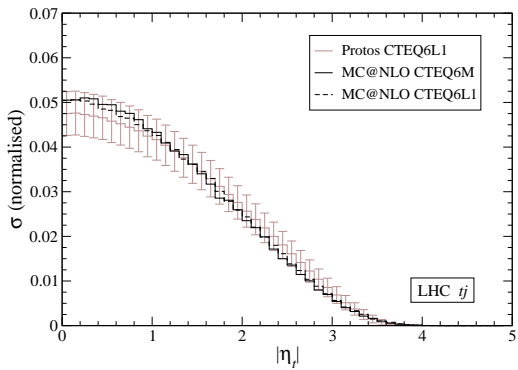}
\caption{Normalised LO rapidity distribution for SM $t$-channel single top production at LHC, including the statistical uncertainties for 10 fb$^{-1}$. The NLO distributions for two different PDFs are also shown.}
\label{fig:rapcomp}
\end{center}
\end{figure}

For our fits we use the LO rapidity distributions for initial $d,s,b$ quarks. It is expected that NLO corrections do not significantly change the shape of these distributions and, in any case, theoretical uncertainties are much smaller than the statistical ones. We illustrate this in Fig.~\ref{fig:rapcomp} for $t$-channel production at LHC, which is the process in which statistics are better. The gray distribution corresponds to the LO production in the SM, {\em i.e.} from an initial $b$ state, with the error bars indicating the statistical uncertainty for a luminosity of 10 fb$^{-1}$. We note that the uncertainties in the different rapidity bins are obtained from the total statistical error for $t$-channel production in Eqs.~(\ref{ec:statLHC}), which is determined by the total number of signal events, and the SM rapidity distribution.
The black solid line is the normalised distribution at NLO, obtained with {\tt MC@NLO}~\cite{Frixione:2002ik}, using CTEQ6M PDFs, while the dashed line, shown for comparison, corresponds to CTEQ6L1 PDFs. It is also apparent that systematic uncertainties on the rapidity distributions will only be relevant for much higher integrated luminosities. 

\begin{figure}[htb]
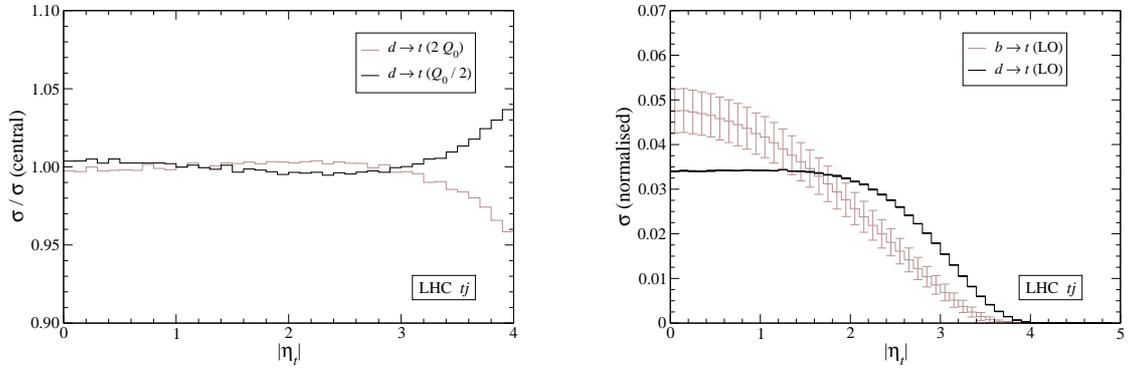
 
\begin{center}
\begin{tabular}{ccc}
\epsfig{file=Figs/fig14a.eps,height=4.8cm,clip=} & \quad \quad & 
\epsfig{file=Figs/fig14b.eps,height=4.8cm,clip=}
\end{tabular}
\caption{Left: ratio of normalised rapidity distributions for the $d \to t$ $t$-channel process using several factorisation choices (see the text). Right: comparison of the $d \to t$ distributions with three scale choices with the one for $b \to t$, including the statistical uncertainties for 10 fb$^{-1}$ for the latter.}
\label{fig:rapcomp2}
\end{center}
\end{figure}

We have also estimated the uncertainty on the rapidity distribution for $d \to t$ by varying the factorisation scale between $Q=2Q_0$ and $Q=Q_0/2$, where $Q_0^2 = -p_W^2$ for the light quark line and $Q_0^2 = -p_W^2 + m_t^2$ for the top quark one corresponds to our central factorisation scale choice. The ratio between the resulting normalised distributions  and the one for $Q=Q_0$ is presented in Fig.~\ref{fig:rapcomp2} (left). 
For the region of interest $|\eta| \lesssim 3.5$ in which most events concentrate the variations are negligible. This is also shown clearly in Fig.~\ref{fig:rapcomp2} (right) where we plot the central ($Q=Q_0$) distribution and its variation (tiny error bars) for $d \to t$, compared to the distribution for $b \to t$ and its statistical error.
On the other hand, PDF uncertainties will likely be under good control, because copious Drell-Yan $W$ production will be used to determine them at these scales by using a $W$ rapidity analysis or an equivalent one.

\begin{figure}[htb]
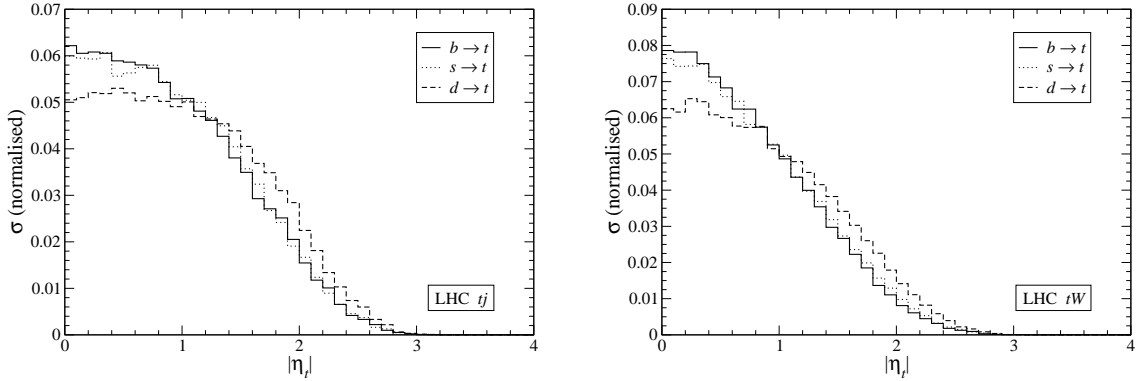

\begin{center}
\begin{tabular}{ccc}
\epsfig{file=Figs/fig12a.eps,height=5cm,clip=} & \quad &
\epsfig{file=Figs/fig12b.eps,height=5cm,clip=}
\end{tabular}
\caption{Normalised rapidity distributions for $t$-channel and $tW$ single top production at LHC after detector simulation}
\label{fig:rapLHC-A}
\end{center}
\end{figure}

Detector and reconstruction effects degrade the top rapidity distributions although they do not wipe out the differences between them.
In order to show this, we have performed a fast simulation of $t$-channel and $tW$ samples for each initial state flavour $d,s,b$ using {\tt Protos} for the event generation (including the top, and $W$ boson spin as well as finite width effects), {\tt Pythia}~\cite{Sjostrand:2006za} for hadronisation and {\tt AcerDet}~\cite{RichterWas:2002ch} for the detector simulation. A $b$ tagging is applied with an efficiency of 60\%, which corresponds to a 15\% mistagging rate for charm jets and 1.1\% for light quark jets. Our selection criteria for the samples are:
\begin{itemize}
\item $t$-channel: one charged lepton with transverse momentum $p_T > 25$ GeV; one $b$-tagged jet with $p_T > 30$ GeV; missing energy $\ptmiss > 25$ GeV.
\item $tW$: one charged lepton with transverse momentum $p_T > 25$ GeV and no other lepton above 10 GeV; one $b$-tagged jet with $p_T > 30$ GeV and two untagged jets also with $p_T > 30$ GeV; missing energy $\ptmiss > 25$ GeV.
\end{itemize}
A very simple reconstruction of the top quark is performed in each case (see Ref.~\cite{Aad:2009wy} for more optimised methods):
\begin{itemize}
\item $t$-channel: the $W$ boson momentum is reconstructed in the usual way taking the neutrino transverse momentum as missing energy, requiring $(p_\ell + p_\nu)^2 = M_W^2$ and choosing for the longitudinal momentum the solution giving a best reconstructed top mass.
\item $tW$: the $W$ boson momentum is reconstructed in the same way but choosing the neutrino momentum solution with smaller longitudinal momentum. Events are accepted only if the invariant mass $m_{\ell \nu b}$ is closer to $m_t$ than $m_{jjb}$, {\em i.e.} if the event is consistent with a semileptonic top decay and a hadronic $W$ decay.
\end{itemize}
The normalised distributions for the positive charge samples are presented in Fig.~\ref{fig:rapLHC-A}.%
It is seen that, even with these non-optimised reconstruction techniques, a good share of the differences existing at parton level between $d$ and $s,b$ initial states are kept. More elaborate reconstruction methods and the application of quality cuts will surely improve the discriminating power, but this analysis falls beyond the scope of this paper.

The background subtraction to isolate the single top signal seems feasible as well. The largest background, $t \bar t$, is charge-symmetric and one can imagine to use the subsample with negatively charged leptons (in which the rapidity is a poor discriminant) to achieve a better background normalisation. Subsequently, this information can be used in the positive charge subample to perform the background subtraction with better precision. In any case, the measurement of rapidity distributions, as any other precision analysis, is a demanding task from the experimental point of view.

\section{Improving constraints with single top rapidity}
\label{sec:3b}

In our fits for the top rapidity analyses we deliberately use the parton-level distributions, in order to show the full potential of this variable to improve the limits on $\vtd$. Of course, as we have indicated in the previous section, this distribution will have to be measured by reconstructing the (single) top quark event candidates, and performing a proper background subtraction.

We obtain our combined limits on $(\vtd,\vts,\vtb)$ by using {\tt TopFit} \cite{AguilarSaavedra:2006fy} extended with the relevant observables for the fit: the single top cross sections in Eqs.~(\ref{ec:xsec-lhc}), (\ref{ec:xsec-tev}), the ratio $R$ and the rapidity distributions in Figs.~\ref{fig:rapLHC} and~\ref{fig:rapTEV}. We generate random points in the $(\vtd,\vts,\vtb)$ parameter space with a flat probability distribution in the range $[0,1.5]$ and use the acceptance-rejection method to obtain a sample distributed according to the combined $\chi^2$ of the observables considered. The limits presented are $1\sigma$ regions with a boundary of constant $\chi^2$ containing 68.26\% of the points accepted.

For the total cross sections and $R$ we perform the fit summing in quadrature the
experimental statistical and systematic uncertainties, and the theoretical one in the former case. In the absence of real data, we take the SM prediction as the ``experimental'' measurement. For the rapidity distributions the analysis is slightly more involved.
In order to construct independent observables, uncorrelated with the total cross sections, we normalise in each case the ``theoretical'' distribution (whose shape and normalisation both depend on $\vtd$, $\vts$ and $\vtb$, taken as free parameters for the fit) to the ``experimental'' one (corresponding to the SM prediction) and sum the $\chi^2$ obtained for each bin.
We do not include any detector effects in the shapes, which could be taken into account by using template methods or correction functions. On the other hand, the statistical errors in the rapidity bins are determined by the total statistical uncertainty in Eqs.~(\ref{ec:statLHC}), obtained with a detailed simulation, and the expected SM distributions.
Bins in the ``experimental'' distribution are required to have at least 5 events, otherwise they are grouped. As we have mentioned before,
in the computation of the $\chi^2$ we only take into account the statistical uncertainty. This seems to be a good first approximation, since (i) some of the systematic uncertainties, for example from the luminosity, only affect the global normalisation, and several other ones should have little dependence on the rapidity of the reconstructed top quark; (ii) rapidity distributions for SM backgrounds are expected to be measured with very good accuracy and well understood, {\em e.g.} in order to determine the quark PDFs from $W$ and $Z$ production, and the associated errors are expected to be smaller than the statistical ones, shown in Fig.~\ref{fig:rapcomp} for the best case. Anyway, systematic uncertainties on the rapidity distributions could be straightforwardly included in {\tt TopFit} for future more detailed analyses.

Finally, it is worth commenting that the differences in top rapidity distributions translate into different pseudo-rapidity spectra for the charged leptons and $b$ quarks resulting from top decay. From the experimental point of view the latter are easier to measure, especially at lower luminosities, because they do not require the reconstruction of the missing neutrino momentum. However, in some cases a large extent of the information from the top rapidity is lost due to spin effects. For example, in $t$-channel production the top quarks are produced with a polarisation $P \simeq -0.9$ in the helicity basis and charged leptons are preferrably emitted in the opposite direction to the top momentum. Therefore, the boost along the initial $d$ quark direction is smaller for the charged lepton, and the differences among $d$, $s$ and $b$ flavours are smeared. On the other hand, for $tW$ production the top polarisation is smaller and the charged lepton pseudo-rapidity spectra show important differences for initial $d$, $s$ and $b$ quarks. These issues will be studied in more detail elsewhere.

\subsection{Limits at LHC}

\begin{figure}[htb]
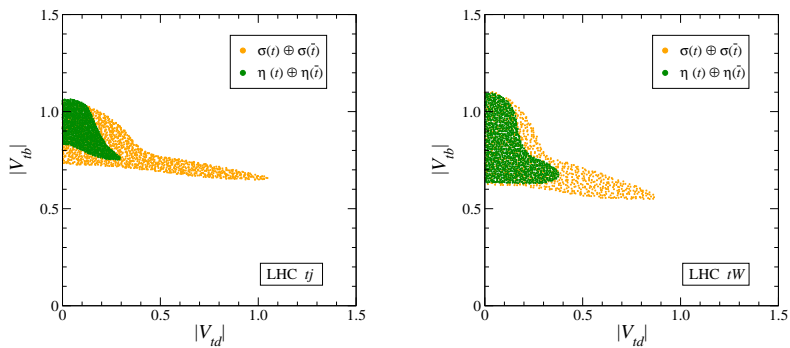

\begin{center}
\begin{tabular}{ccc}
\epsfig{file=Figs/fig8a.eps,height=4.5cm,clip=} & \quad &
\epsfig{file=Figs/fig8b.eps,height=4.5cm,clip=}
\end{tabular}
\caption{Projection on the $(\vtd,\vtb)$ plane of the combined limits from $t$-channel and $tW$ production at LHC, without and with the inclusion of the single top rapidity.}
\label{fig:lim-LHC1}
\end{center}
\end{figure}

The excellent statistics for $t$-channel and $tW$ production at LHC allows to constrain $\vtd$ only using either of these processes. For illustration, we show in Fig.~\ref{fig:lim-LHC1} the projections of the limits on the $(\vtd,\vtb)$ plane, using the cross section measurements $\sigma(t)$ and $\sigma(\bar t)$ and also including the rapidities $\eta(t)$, $\eta(\bar t)$. In the former case $\vtd$ is practically unconstrained, while the cross sections and rapidity distributions set the bound $|\vtd| \lesssim 0.4$. For $\vts$ the limits are practically unchanged.

\begin{figure}[htb]
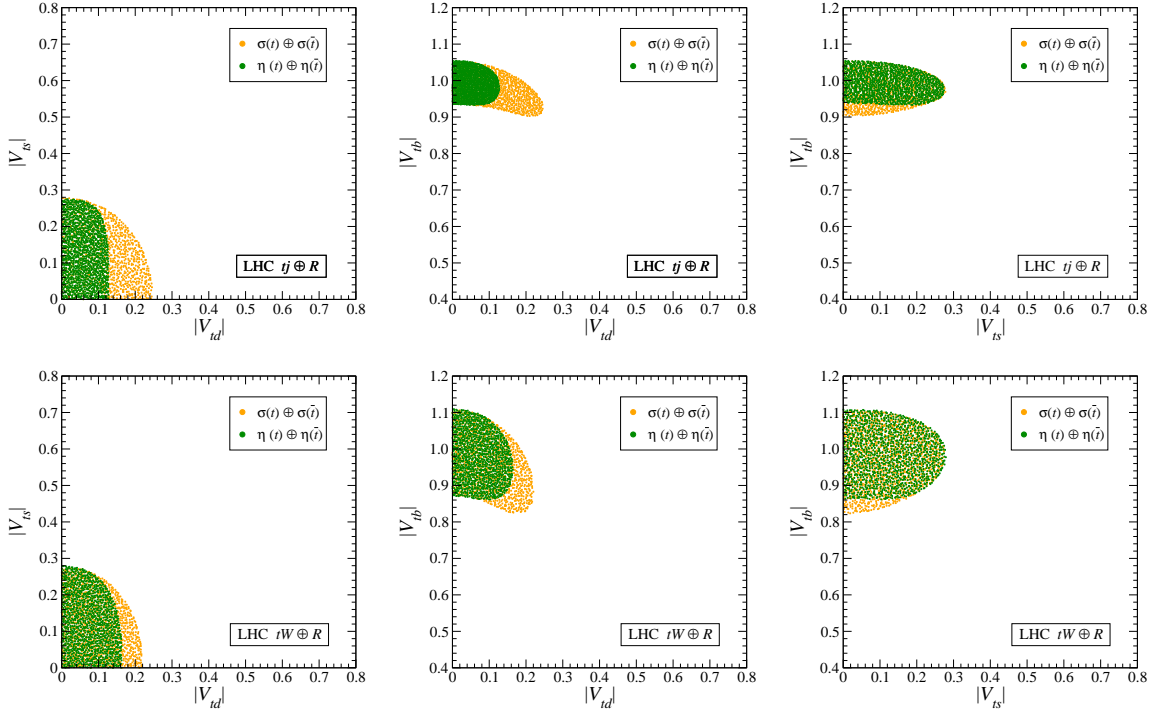

\begin{center}
\begin{tabular}{ccc}
\epsfig{file=Figs/fig9a.eps,height=4.5cm,clip=} &
\epsfig{file=Figs/fig9b.eps,height=4.5cm,clip=} &
\epsfig{file=Figs/fig9c.eps,height=4.5cm,clip=} \\[2mm]
\epsfig{file=Figs/fig9d.eps,height=4.5cm,clip=} &
\epsfig{file=Figs/fig9e.eps,height=4.5cm,clip=} &
\epsfig{file=Figs/fig9f.eps,height=4.5cm,clip=}
\end{tabular}
\caption{Projections on the $(\vtd,\vts)$, $(\vtd,\vtb)$ and $(\vts,\vtb)$ planes of the combined limits from $t$-channel (up) and $tW$ production (down) in combination with $R$ at LHC.}
\label{fig:lim-LHC2}
\end{center}
\end{figure}

\begin{figure}[htb]
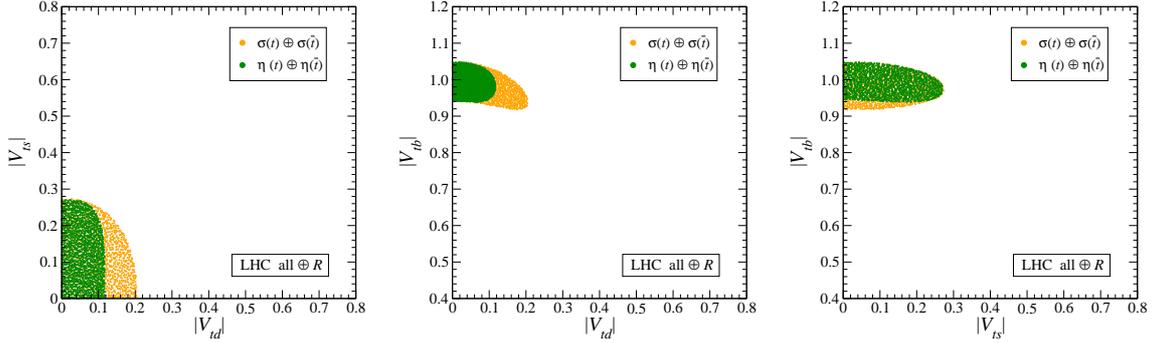

\begin{center}
\begin{tabular}{ccc}
\epsfig{file=Figs/fig10a.eps,height=4.5cm,clip=} &
\epsfig{file=Figs/fig10b.eps,height=4.5cm,clip=} &
\epsfig{file=Figs/fig10c.eps,height=4.5cm,clip=}
\end{tabular}
\caption{Projections on the $(\vtd,\vts)$, $(\vtd,\vtb)$ and $(\vts,\vtb)$ planes of the combined limits  from all single top channels and $R$ at LHC.}
\label{fig:lim-LHC3}
\end{center}
\end{figure}

We present in Fig.~\ref{fig:lim-LHC2} the result of the global fits including either $t$-channel (up) or $tW$ production (down) and $R$. We point out that significant constraints on the $(\vtd,\vts,\vtb)$ parameter space can be set by using
only one of these single top processes in combination with the $R$ measurement from $t \bar t$ production.
The limits including both processes as well as the $s$-channel cross section (which has a negligible impact) are given in Fig.~\ref{fig:lim-LHC3}. We observe that
the global limits on $\vtd$ are reduced by a factor of two with the rapidity distribution analysis, from $|\vtd| \leq 0.21$ to $|\vtd| \leq 0.12$. The limits on $\vtb$ are also reduced, from $0.92 \leq |\vtb| \leq 1.05$ to $0.94 \leq |\vtb| \leq 1.05$, whereas the bound $|\vts| \leq 0.27$ is not significantly affected.\footnote{We remind the reader that these intervals are not $1\sigma$ limits on the individual mixings but the range of variation of the parameters in the $1\sigma$ volume.} These figures may be degraded with detector effects, but an improvement is expected in any case when the top rapidity distribution is used.

\subsection{Limits at Tevatron}

\begin{figure}[htb]
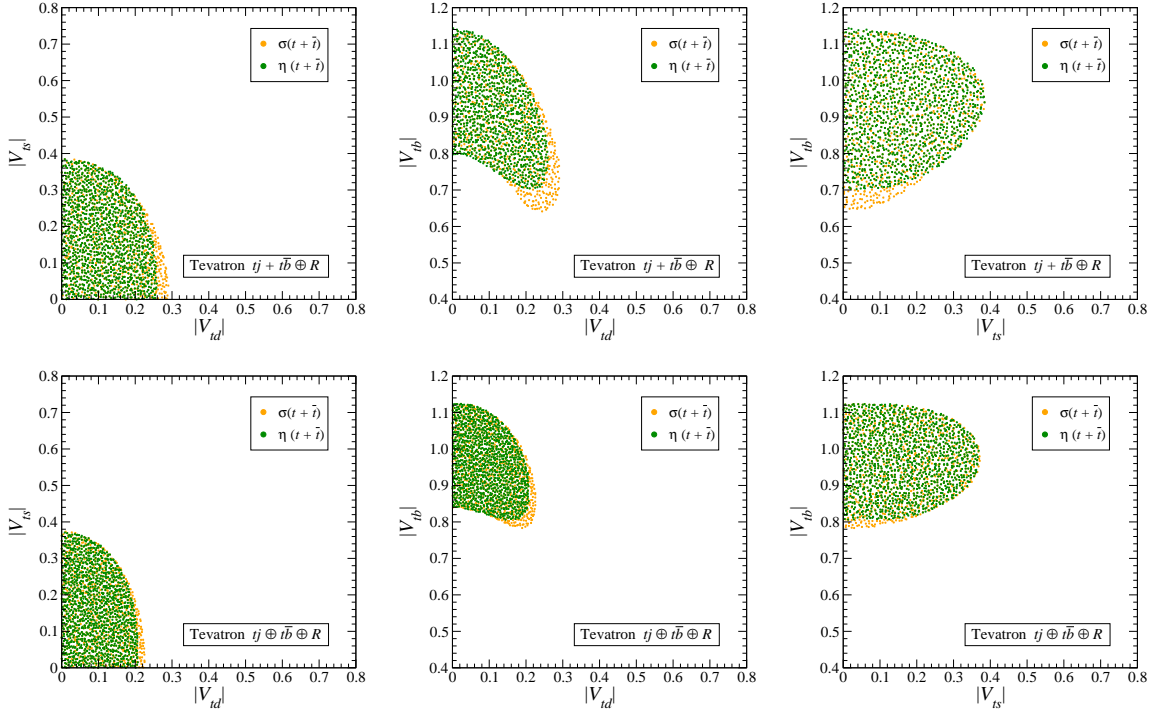

\begin{center}
\begin{tabular}{ccc}
\epsfig{file=Figs/fig11a.eps,height=4.5cm,clip=} &
\epsfig{file=Figs/fig11b.eps,height=4.5cm,clip=} &
\epsfig{file=Figs/fig11c.eps,height=4.5cm,clip=} \\[2mm]
\epsfig{file=Figs/fig11d.eps,height=4.5cm,clip=} &
\epsfig{file=Figs/fig11e.eps,height=4.5cm,clip=} &
\epsfig{file=Figs/fig11f.eps,height=4.5cm,clip=} 
\end{tabular}
\caption{Projections on the $(\vtd,\vts)$, $(\vtd,\vtb)$ and $(\vts,\vtb)$ planes of the combined limits from single top production and $R$ at Tevatron.}
\label{fig:lim-TEV}
\end{center}
\end{figure}

Despite the experimental challenges for single top observation at Tevatron~\cite{Abazov:2009ii,Aaltonen:2009jj}, a future analysis of the top rapidity distributions might improve the global fits. This measurement obviously demands significant statistics but does not require the separate identification of the $t$- and $s$-channel processes. We show in Fig.~\ref{fig:lim-TEV} the combined limits using the sum of $s+t$-channel cross sections (up) or combining their separate measurements (down) as two extreme cases. In both cases we include the ratio $R$ as well. The improvement brought by the top rapidity distribution is not as clear as for LHC, but the results of our fit suggest that the limits on $\vtd$, $\vtb$ might both be reduced:
\begin{itemize}
\item[(i)] For an inclusive measurement (upper plots) the reduction is of 10\% in the $\vtd$, $\vtb$ bounds, resulting in $|\vtd| \leq 0.26$, $|\vts| \leq 0.38$, $0.70 \leq \vtb \leq 1.14$.
\item[(ii)] If the $s$- and $t$-channel cross sections are measured independently, the reduction is around 8\%, resulting in $|\vtd| \leq 0.21$, $|\vts| \leq 0.37$, $0.80 \leq \vtb \leq 1.12$.
\end{itemize}
Therefore,
a detailed analysis would be welcome if sufficient data is collected. We also note that the estimated Tevatron limit on $\vtd$ is as good as the one expected for LHC if the top rapidity distribution is not used for the latter (see the previous subsection).

\section{Summary}
\label{sec:4}

Single top measurements at Tevatron and LHC are essential for the determination of the top quark charged current interactions, in particular the CKM matrix elements $\vtd$, $\vts$ and $\vtb$. In this paper we have pointed out the important role of the single top rapidity distribution in order to discriminate the production of top quarks from $d$ initial states against $s$ and $b$. This is important not only for the experimental determination of $\vtd$ but to improve the limits on $\vtb$: both of them, as well as $\vts$, must be obtained from a global fit to several observables, and improving the constraints on $\vtd$ also tightens the limits on $\vtb$. 
We have shown that, with its excellent statistics, LHC can take advantage of this distribution to improve the bound on $\vtd$ by a factor of two. With a luminosity of 10 \fbin\ at 14 TeV, the limits 
$|\vtd| \leq 0.12$, $|\vts| \leq 0.27$, $0.94 \leq |\vtb| \leq 1.05$ could be achieved.
At Tevatron with a luminosity of 12 \fbin\, the corresponding limits would be
$|\vtd| \leq 0.21$, $|\vts| \leq 0.37$, $0.80 \leq \vtb \leq 1.12$
if $s$- and $t$-channel cross sections can be measured independently with a good precision. Otherwise,
the limits would be less stringent, $|\vtd| \leq 0.26$, $|\vts| \leq 0.38$, $0.70 \leq \vtb \leq 1.14$.
It is also important to remark here that LHC will improve significantly the Tevatron limit on $\vtd$ only if the rapidity distribution is used or if experimental uncertainties are greatly reduced with respect to present expectations~\cite{Aad:2009wy}.

Our analysis highlights the importance of separating $t$ and $\bar t$ events in single top production at LHC. For total cross sections, the different functional dependence on $\vtd$, $\vts$ and $\vtb$ can be exploited to set more stringent bounds on them. For rapidity measurements the separation is even more important, because the differences between initial $d$ and $s,b$ flavours are much more pronounced for final state $t$ quarks than for antiquarks. We also emphasise the importance of
measuring $R = \text{Br}(t \to Wb)/\text{Br}(t \to Wq)$ in $t \bar t$ production. At LHC, the $s$-channel cross section determination will have large experimental errors, hence the measurement of $R$ is essential to set limits on the top CKM matrix elements in combination with the $t$-channel and/or $tW$ cross sections. At Tevatron, this observable is also required if the $s$- and $t$-channel cross sections are not measured separately, and improves the limits in any case.

Finally, the results presented here make apparent an obvious fact: the total cross sections are not the only observables sensitive to $\vtd$, $\vts$ and $\vtb$ in single top production processes. Indeed, the determination of these mixings at Tevatron and LHC will be better achieved by using template methods and performing a global fit, including not only the cross sections but also the rapidity and other distributions possibly sensitive to the top mixing parameters. This work is left for future experimental studies.

\section*{Acknowledgements}

We thank B. Casal and A. Ruiz for useful discussions and correspondence. This work has been partially supported by CRUP (Ac\c{c}\~ao integrada Ref. E 2/09),
FCT (project CERN/FP/83588/2008),
MICINN (FPA2006-05294, FPA2010-17915 and HP2008-0039),
Junta de Andaluc\'{\i}a (FQM 101 and FQM 437),
and by the European Community's Marie-Curie Research Training
Network under contract MRTN-CT-2006-035505 ``Tools and Precision
Calculations for Physics Discoveries at Colliders''.
The work of J.A.A.S. has been supported by a MICINN Ram\'on y Cajal contract.

\end{document}